\documentstyle[aps,multicol,graphics,prb]{revtex}
\begin{document}
\draft
\title{Analysis of current-voltage characteristics of
two-dimensional superconductors:
finite-size scaling behavior in the vicinity of
the Kosterlitz Thouless transition}

\author{Kateryna Medvedyeva, Beom Jun Kim, and  Petter Minnhagen}
\address {Department of Theoretical Physics, Ume{\aa} University,
901 87 Ume{\aa}, Sweden}
\preprint{\today}
\maketitle
\thispagestyle{empty}

\begin{abstract}
It has been suggested [Pierson {\it et al.}, Phys. Rev. B {\bf 60}, 1309
(1999); Ammirata {\it et al.}, Physica C {\bf 313}, 225 (1999)] that for
two-dimensional (2D) superconductors there exists a phase transition with the
dynamic critical exponent $z\approx 5.6$.  We perform simulations for the 2D
resistively-shunted-junction model and compare the results with the
experimental data in Repaci {\it et al.} obtained for an ultrathin YBCO sample
[Phys. Rev. B {\bf 54}, R9674 (1996)]. We then use a different method of
analyzing dynamic scaling than in Pierson {\it et al.}, and conclude that both
the simulations and the experiments are consistent with a conventional
Kosterlitz Thouless (KT) transition in the thermodynamic limit for which $z=2$.
For finite systems, however, we find both in simulations and experiments that
the change in the current-voltage ($I$-$V$) characteristics caused by the
finite size shows a scaling property with an exponent $\alpha\approx 1/6$,
seemingly suggesting a vanishing resistance at a temperature for which
$z=\alpha^{-1}$.  It is pointed out that the dynamic critical exponent found in
Pierson {\it et al.} corresponds to the exponent $\alpha^{-1}$.  It is
emphasized that this scaling property does not represent any true phase
transition since in reality the resistance vanishes only at zero temperature.
Nevertheless, the observed scaling behavior associated with $\alpha \approx
1/6$ appears to be a common and intriguing feature for the finite size caused
change in the $I$-$V$ characteristics around the KT transition.  
\end{abstract}


\begin{multicols}{2}
\section{Introduction} \label{sec:intro} 
The transition from the superconducting to the resistive state is of the
Kosterlitz Thouless (KT) type for an ideal two-dimensional (2D) superconductor
in zero magnetic field.~\cite{minnhagen:rev} Below the KT transition the
current-voltage ($I$-$V$) characteristics is nonlinear, $V\propto I^a$ (or
equivalently, $E \propto J^a$ with the electric field $E$ and the current
density $J$), characterized by the exponent $a$.  According to the conventional
extension of the KT theory, $a=3$ precisely at the transition and $a$ becomes
larger as the temperature is
decreased.~\cite{minnhagen:rev,ahns,kadin,minnhagen:lett,beom:big} The scaling
argument by Fisher, Fisher, and Huse (FFH) in Ref.~\onlinecite{ffh} leads to
the connection $a=1+z$ at the KT transition from which the value $z=2$ is
deduced for the dynamic critical exponent $z$.~\cite{dorsey} These predictions
of $z=2$ and $a=3$ at the KT transition have been confirmed through numerical
simulations for the lattice Coulomb gas with Monte Carlo (MC)
dynamics,~\cite{teitel} the Langevin-type molecular dynamics of Coulomb gas
particles,~\cite{holmlund} and the 2D $XY$ model with 
resistively-shunted-junction (RSJ) dynamics.~\cite{minnhagen:lett,beom:big,melwyn} The value $a=3$
has also been found to be consistent with numerous
measurements.~\cite{minnhagen:rev} 

From the above perspective, the recent claim by Pierson {\it et al.} in
Ref.~\onlinecite{pierson} that $z$ has a much larger value at the resistive
transition is very intriguing and is in apparent contradiction to the
established conventional view.  More specifically, Pierson {\it et al.} in
Ref.~\onlinecite{pierson} have applied a dynamic FFH scaling approach (we call it
Pierson scaling in this work; see Sec.~\ref{sec:scaling}) to existing experimental
data and obtained  $z\approx 6$ in many 2D systems including
theoretical models like the lattice Coulomb gas (Ref.~\onlinecite{teitel}),
$^4$He films, and an ultrathin YBCO sample (Repaci {\it et al.} in
Ref.~\onlinecite{repaci}).  We in our analysis focus on the Repaci {\it et al.}
data for which a particularly good scaling collapse was obtained by
Pierson {\it et al.}

Our approach is, as directly as possible, to compare simulation data for the 2D
RSJ model obtained in this work with the experimental data obtained
for a ultrathin YBCO film by Repaci {\it et al.} in Ref.~\onlinecite{repaci}.  (For
brevity, we from now on call the former the RSJ data and the latter the Repaci data,
respectively.) In case of the RSJ data we conclude that the established
conventional view holds with $z=2$ at the KT transition.  We then compare the
RSJ data with the Repaci data and show that the latter exhibit a similar
evidence of $z=2$ at the KT transition.  Both the RSJ and the Repaci data show
nonvanishing resistances below the KT transition. In the former case this is due to
finite-size effects and in the latter it is due to some length scale which
introduces an effective cutoff in the logarithmic vortex interaction, e.g., the
finite size of the sample, the perpendicular penetration depth, or a residual weak
perpendicular magnetic field (see Ref.~\onlinecite{minnhagen:rev} for details);
We in the following use the term finite-size effects to cover all these
possibilities.  In the original analysis of the data in
Ref.~\onlinecite{repaci} the authors attributed the finite resistance below the
KT transition to such finite-size effects, which  is in accord with the
striking similarity we find to the RSJ data.

We also find that the Pierson scaling can be made to work with $z\approx 6$ for
the RSJ data which is closely the same $z$ as found by Pierson {\it et al.} for 
the Repaci data.  In addition, we find, by analyzing the
resistive tails of both data sets through the use of a new scaling method,
that the finite-size caused resistance tails show strikingly similar
scaling behaviors with an exponent $\alpha\approx 1/6$.  However, our
conclusion is that this scaling does not signal any real resistive transition,
but reflects finite-size effects which will eventually disappear in
the thermodynamic limit.

Strachan {\it et al.} (Ref.~\onlinecite{strachan}) have, from a thorough analysis
of the Repaci data, shown that the Pierson method is too flexible to
allow any firm
conclusion and that the apparent vanishing of the resistance is really a
consequence of the finite-voltage resolution threshold in the experiment.
Somewhat related conclusions have been reached by Brown in Ref.~\onlinecite{brown}
in the context of the vortex glass transition. This is also in accord with our
findings.  Nevertheless our conclusion is that the Pierson scaling result
is, at least partially, a reflection of an actual scaling like behavior
associated with the finite-size caused resistance tail in the vicinity of the
KT transition. However, since the resistance tail vanishes only at zero
temperature for a finite system, we conclude that there exists no resistive
transition at a finite temperature and hence no new transition. A possible
underlying reason for the scaling behavior is still an enigma which remains to
be resolved in the future.

The present paper is organized as follows: In Sec.~\ref{sec:scaling} the
scaling methods used to analyze the data are presented, and in
Sec.~\ref{sec:rsj} we describe and analyze the RSJ data we obtain
from the
simulations. The Repaci's YBCO data is reanalyzed and compared with the RSJ
data in Sec.~\ref{sec:repaci}.  Finally we summarize and discuss our findings
in Sec.~\ref{sec:conc}.

\section{Scaling Methods} \label{sec:scaling}

The dynamic scaling relation proposed by FFH (Ref.~\onlinecite{ffh}) for a
$d$-dimensional superconductor in the vicinity of  the transition is given
by~\cite{foot:EJ} 
\begin{equation} \label{eq:ffh}
E=J{\xi}^{d-2-z}\chi_\pm(J\xi^{d-1}/T) , 
\end{equation} 
where $E$, $J$, $\xi$,
$z$, and $T$ are the electric field, the applied current density, the correlation
length, the dynamic critical exponent, and the temperature, respectively, and
$\chi_\pm$ denotes the scaling function above ($+$) and below ($-$) the
transition.  It should be kept in mind that Eq.~(\ref{eq:ffh}) can be directly
used only in a system whose linear size $L$ is much larger than $\xi$ ($L \gg \xi$).

Specializing to two dimensions, $d=2$, Eq.~(\ref{eq:ffh}) can be recast into the
form 
\begin{equation} \label{eq:p1}
\frac{E}{J}=\xi^{-z}\chi_\pm(J\xi/T),
\end{equation}
which is the starting point in Pierson {\it et al.}~\cite{pierson} It should be
noted that the above FFH dynamic scaling form (\ref{eq:p1}) is valid only for a
phase transition when $\xi$ is finite both above and below the transition
temperature.
Accordingly,  one should note that Eq.~(\ref{eq:p1}) is not
compatible with a system undergoing a KT transition because the whole
low-temperature phase is "quasi" critical with $\xi=\infty$ and is
characterized by a line of fixed points.~\cite{minnhagen:rev,footnew1}
This means that
only the scaling function $\chi_+$  above the KT transition has any
clear justification.~\cite{foot}

In the present paper we study in Sec.~\ref{sec:rsj} the 2D RSJ model with a
finite size $L$, which, instead of $\xi$, serves as the large length scale
cutoff at and below the KT transition.  At the KT transition the finite-size
scaling form deduced from Eq.~(\ref{eq:p1}) by substituting $\xi$ by $L$ takes
the form:~\cite{ffh,teitel}
\begin{equation} \label{eq:ktscale}
\frac{E}{J}=L^{-z}f(JL) , 
\end{equation}
where the scaling function $f(x)$ satisfies $f(0)=$const. and $f(x)\propto
x^z$ for large $x$, corresponding to the finite-size induced resistance
$R\propto L^{-z}$ without external current and $E \propto J^{1+z}$ in the
large-current limit, respectively.  The conventional view certifies that the
nonlinear $I$-$V$ exponent $a$ in $E\propto J^a$, is equal to 3 precisely at
the KT transition~\cite{ahns,minnhagen:lett} from which $z=2$ follows by the
scaling relation $a=z+1$.~\cite{dorsey}

Since the low-temperature phase in the KT scenario is described by a line of
fixed points, each temperature $T$ below the KT transition is characterized by
its own scaling function $h_T(x)$:
\begin{equation} \label{eq:ourscale}
\frac{E}{JR}=h_T \bigglb( JLg_L(T) \biggrb) , 
\end{equation}
where $R = \lim_{J \rightarrow 0}(E/J) \propto L^{-z(T)}$ [compare
with Eq.~(\ref{eq:ktscale})], $h_T(0)=1$, $h(x)\propto x^{z(T)}$ for large $x$,
and $g_L(T)$ is a function of at most $T$ and $L$ such that a finite
limit function $g_\infty(T)$ exists in the large-$L$ limit.  

The temperature-dependent dynamic critical exponent $z(T)$ can be obtained from
a simple scaling argument (See Ref.~\onlinecite{melwyn} for more details): The
characteristic time $\tau$, which is related with $R$ by $R \sim 1/\tau$, is inversely
proportional to the rate of unbinding vortex pairs over the boundary,
and is given by the escape-over-barrier rate
\begin{equation}  \label{eq:rate} 
\frac{1}{\tau} \propto n L^2 \exp\left(-\frac{1}{\epsilon T^{CG}}\ln L\right), 
\end{equation}
where $n$ is the  average number density of vortices and the escape barrier for
a pair to unbind over the boundary is determined by the effective logarithmic
vortex interaction $(1/\epsilon)\ln L$.  Here $\epsilon$ and $T^{CG}$ are the
dielectric constant and the Coulomb gas temperature, respectively (see
Ref.~\onlinecite{minnhagen:rev}), and both can be calculated in equilibrium.
From the definition of $z(T)$ ($\tau \sim L^z$) and Eq.~(\ref{eq:rate}) we
find~\cite{minnhagen:lett} 
\begin{equation} \label{eq:z}
z(T) =\frac{1}{\epsilon T^{CG} }-2.
\end{equation}
Furthermore $T^{CG}\propto T/\rho_0(T)$, where $\rho_0(T)$ is the density
of Cooper pairs which decreases with increasing  $T$, resulting in $T^{CG}$
monotonically decreasing to zero as $T \rightarrow 0$.~\cite{minnhagen:rev} On the other hand,
$\epsilon (T)\geq 1$ and remains close to unity in most of the low-temperature
phase. It then follows from Eq.~(\ref{eq:z}) that $z(T)$ increases
monotonically as $T$ is lowered, reaching $z(T)=\infty$ at $T=0$.  One may note
that since the KT transition is characterized by the condition $1/\epsilon
T^{CG}=4$ (Refs.~\onlinecite{minnhagen:rev} and \onlinecite{kosterlitz-nelson})
Eq.~(\ref{eq:z}) results in the conventional established value $z=2$ at the KT
transition.  

Since $E/JR \propto x^{z(T)}$ for large $x$ [see Eq.~(\ref{eq:ourscale})] and
$z(T)$ depends on $T$ [see Eq.~(\ref{eq:z})], $E/JR$ cannot be described by
a $T$-independent scaling function.  On the other hand, when the scaling
variable $x = JLg_L(T)$ has small values, the possibility that $E/JR$ is
described by a $T$-independent scaling function with the dominant $T$-dependence
absorbed into $x$, cannot {\it a priori} be excluded.  We find from our simulations
of the 2D RSJ model that this indeed appears to be the case and one of the main
points in the present paper is the apparent existence of the scaling form
\begin{equation} \label{eq:newscale}
\frac{E}{JR}=h\bigglb(JLg_L(T)\biggrb)
\end{equation}
for smaller values of $JLg_L(T)$ [compare with Eq.~(\ref{eq:ourscale})].  What
can be concluded about the function $g_L(T)$ on general grounds?  First we note
that the resistance $R\propto n(T)L^{-z(T)}$ for a {\em finite system size}
vanishes only at zero temperature but has a finite value at any nonzero
temperature. This follows since $z(T)<\infty$ at any nonzero $T$ and $n(T)$ is
of the thermal activation form, i.e., $n(T)\propto e^{-|{\rm const.}|/T^{CG}}$
(Ref.~\onlinecite{minnhagen:rev}), which is finite at any nonzero temperature.
A finite $R$ by Eq.~(\ref{eq:ourscale}) implies that $0<g_L(T)<\infty$ and our
simulations of the 2D RSJ model are entirely consistent with this expectation.
If, on the other hand, $R$ would vanish at a finite $T$ this would have
implied a diverging $g_L(T)$ at this temperature.

The finite-size scaling we are discussing here is on the face of it entirely
different from the scaling form given by Eq.~(\ref{eq:p1}), which
is correct only in the thermodynamic limit $\xi / L \rightarrow
0$ and  constitutes the basis and theoretical underpinning for the Pierson
method in Ref.~\onlinecite{pierson}.  We again emphasize that this
thermodynamic scaling is only justified as the KT transition is approached from
above and has no bearing on the nonzero resistance below the KT transition
caused by the finite size of the sample. However, from the practical standpoint
of analyzing procedures the Pierson method has a connection to our finite-size
scaling form $E/JR=h(JLg_L(T))$:
The former essentially amounts to assuming that $\xi$ in Eq.~(\ref{eq:p1}) is
proportional to $R^{-\alpha}$, where $\alpha$ is a $T$-independent constant,
leading to the scaling form $E/JR=\tilde{h}(JR^{-\alpha})$ [we have in
this consideration ignored an additional weak unimportant $T$ dependence in
Eq.~(\ref{eq:p1})]. Consequently, if
$g_L(T)=A_LR^{-\alpha}$ where $A_L$ is a constant which may depend on
$L$,
then the Pierson scaling in a practical sense is
of the same form as our finite-size scaling, in spite of the totally
incompatible theoretical ground. So from this point of view the question is to
what extent the relation $g_L(T)=A_LR^{-\alpha}$ is fulfilled. If we allow
$\alpha$ to be a function of both temperature and size then the relation
$g_L(T)=A_LR^{-\alpha_L(T)}$ just serves as a definition of $\alpha_L(T)$
because both $g_L(T)$ and $R(T)$ are positive functions. Thus, in the present
finite-size context and from a practical standpoint, the Pierson scaling
amounts to the statement that $\alpha_L(T)$ extracted from experimental data
appears to be a $T$-independent constant to a good approximation.

The scaling form Eq.~(\ref{eq:ourscale}) implies that $R(T)\propto
[Lg_L(T)]^{-z(T)}$ where the proportionality factor in general depends on
temperature.  Using the definition of $\alpha_L(T)$,
$g_L(T)=A_LR^{-\alpha_L(T)}$, we get the connection
$[Lg_L(T)]^{\alpha_L(T)z(T)}=B_L(T)g_L(T)$, where $B_L(T)$ in general
may depend on both $T$ and $L$, which means that
\begin{equation} \label{eq:g}
\ln g_L(T)=\frac{\alpha_L(T)z(T)}{1-\alpha_L(T)z(T)}\ln \frac{L}{B_L(T)}-\ln B_L(T) .
\end{equation}
The point with this relation is that if the dominant $T$-dependence of
 $R$ comes from the factor $(Lg_L(T))^{-z(T)}$ then the $T$-dependence of
$B_L(T)$ is unimportant.  Eq.~(\ref{eq:g}) then leads to the requirement
$z(T)<1/\alpha_L(T)$ where the extreme condition $z(T)=1/\alpha_L(T)$ for a
$T>0$, if it occurred, would correspond to a diverging $g_L(T)$ and hence to a
vanishing resistance.  This means that if  $\alpha_L(T)$ for a fixed size $L$
to a good approximation was a constant $\alpha$, then the implication would be
that the resistance vanishes at the temperature for which $z(T)=1/\alpha$.  As
discussed above this does not happen within the finite size KT scenario.  But
it could happen that $\alpha_L(T)$ is a constant, i.e., $\alpha_L(T) = \alpha$, 
over a limited temperature region and if  all the available
data was within this region one could be tempted to conclude that such a
transition did really occur at some smaller $T$ outside this region. 
As will be shown in Sec.~\ref{sec:rsj} we do find
such a limited region for the 2D RSJ model.  

\section{Analysis of RSJ data} \label{sec:rsj}
The 2D $L\times L$ $XY$ model on a square
lattice is often used for studies of the
KT transition. It is defined in terms of the Hamiltonian 
\begin{equation} \label{eq:H}
  H=-E_J\sum_{\langle ij\rangle}\cos (\theta_i-\theta_j-{\bf r}_{ij}\cdot{\bf
\Delta}), 
\end{equation}
where $E_J$ is the Josephson coupling strength, the summation is over
nearest-neighbor pairs, $\theta_i$ is the phase angle at site $i$ satisfying
the periodicity: $\theta_i=\theta_{i+L\hat{\bf x}}=\theta_{i+L\hat{\bf y}}$.
Here we use the fluctuating twist boundary condition (FTBC) to allow the phase
difference across the whole system to fluctuate, and include the twist variable
${\bf \Delta} = (\Delta_x, \Delta_y)$ in the Hamiltonian (\ref{eq:H}), where
${\bf r}_{ij}$ is the unit vector from site $i$ to site $j$; ${\bf \Delta}
\equiv 0$ corresponds to the standard periodic boundary condition.  In
particular, FTBC is very efficient when the system is driven by an external
current since the periodicity of phase variables $\theta_i$ is
preserved.~\cite{beom:big} 

In the RSJ dynamics the local current is conserved at every site,
which determines the equations of motion for phase variables:~\cite{beom:big}
\begin{equation} \label{eq:theta}
\dot{\theta}_i=-\sum_j G_{ij}{\sum_k}^\prime
[\sin(\theta_j-\theta_k -{\bf r}_{jk}\cdot{\bf \Delta})+\eta_{jk}] , 
\end{equation}
where the primed summation is over the four nearest neighbors of $j$,
$G_{ij}$ is the lattice Green function for 2D square lattice, 
time $t$ has been normalized in units of $\hbar /2eri_c$
with the critical current $i_c$ and the shunt resistance $r$ 
of a single junction, and $\eta_{jk}$ is the thermal noise current 
in units of $i_c$ on the link from the site $j$ to $k$.
The equations of motion for the twist variables in the presence
of an external current density $J$ in the $x$ direction read
\begin{eqnarray}
\dot\Delta_x & = & \frac{1}{L^2}\sum_{\langle ij\rangle_x}
\sin(\theta_i-\theta_j-\Delta_x)+\eta_{\Delta_x}-J , \label{eq:delta_x} \\
\dot\Delta_y & = & \frac{1}{L^2}\sum_{\langle ij\rangle_y}
\sin(\theta_i-\theta_j-\Delta_y)+\eta_{\Delta_y} , \label{eq:delta_y} 
\end{eqnarray}
which have been obtained from the condition that the net global current across
the system in each direction should vanish (see Refs.~\onlinecite{beom:big} and
\onlinecite{melwyn} for discussions).  Here $\sum_{\langle ij\rangle_x}$
denotes the summation over all bonds in the $x$ direction, $J$ is in units of
$i_c$ (the lattice constant is set to unity).  The thermal noise terms
$\eta_{ij}$ in Eq.~(\ref{eq:theta}), $\eta_{\Delta_x}$ in
Eq.~(\ref{eq:delta_x}), and $\eta_{\Delta_y}$ in  
Eq.~(\ref{eq:delta_y}) obey the conditions $\langle\eta_{ij}\rangle= \langle
\eta_{\Delta_x}\rangle = \langle\eta_{\Delta_y}\rangle =\langle\eta_{ij}
\eta_{\Delta_x}\rangle =\langle\eta_{ij} \eta_{\Delta_y}\rangle
=\langle\eta_{\Delta_x} \eta_{\Delta_y}\rangle =0$,
$\langle\eta_{\Delta_x}(t)\eta_{\Delta_x}(0)\rangle=
\langle\eta_{\Delta_y}(t)\eta_{\Delta_y}(0)\rangle=(2T/L^2)\delta(t)$, and
$\langle\eta_{ij}(t)\eta_{kl}(0)\rangle=2T\delta(t)
(\delta_{ik}\delta_{jl}-\delta_{il}\delta_{jk})$, where $T$ is in units of
$E_J/k_B$. The voltage drop $V$ across the system in the $x$ direction is
written as $V = -L \dot\Delta_x$ in units of $i_c r$, and we measure
time-averaged (denoted by $\langle \cdots \rangle_t$) electric field $E =
\langle V  / L\rangle_t$ to obtain the $I$-$V$ characteristics.  The set of
equations (\ref{eq:theta}), (\ref{eq:delta_x}), and (\ref{eq:delta_y}) are
integrated by using the second order Runge-Kutta-Helfand-Greenside
algorithm~\cite{batrouni} with discrete time step $\Delta t = 0.05$ for $L=4,
6$, and 8, and the electric fields $E$'s are measured from time-averages over
$O(10^6)$ time steps for large currents and $O(10^8)$ steps for small currents.
 
Figure~\ref{fig:rsjall} shows the $I$-$V$ characteristics ($E$ versus $J$) at
various temperatures in log scales for the system size $L=8$.  In the
small-current regime $E$ is shown to approach the Ohmic behavior ($E/J$ =
const.) at any temperature used in the simulations.  Since it is well
established that the 2D $XY$ model on the square lattice has the KT transition
at $T_{KT} \approx 0.892$ (Ref.~\onlinecite{olsson}), the ubiquitous
nonvanishing resistance tails in the small-current regime are clearly due to the
finite-size effects, and in the thermodynamic limit the resistance $R \equiv
\lim_{J \rightarrow 0} (E/J)$ vanishes in the whole low-temperature phase
below $T_{KT}$ giving rise to a nonlinear $I$-$V$ characteristics: 
$E\propto J^{z(T)+1}$ where $z(T)=1/\epsilon T^{CG}-2$ 
can be readily obtained from the equilibrium properties.~\cite{beom:big}
At the KT transition $z(T_{KT})=2$ which corresponds to $E\propto J^3$
(denoted by thick full line in Fig.~\ref{fig:rsjall} 
at $T=0.90$ very close to $T_{KT}$) and crosses over to the Ohmic behavior $E/J =$const. as $J$ is decreased
due to the finite-size effect (thick broken line). 

The nonlinear $I$-$V$
characteristics for the somewhat larger currents is caused by the breaking of
vortex-antivortex pairs due to the finite current. It is this pair breaking
which is described by the exponent $a(T) = z(T)+1$; it continuously
increases from 1 in the high-temperature limit
to infinity at $T=0$ as $T$ is decreased.~\cite{holmlund} 
Consequently one should note
that there is no jump in the exponent describing the current induced
vortex-antivortex pair breaking at the KT transition. We stress this point
because this fact seems sometimes to be misunderstood and to cause confusions.

 Suppose now that the $I$-$V$ characteristics in Fig.~\ref{fig:rsjall} was all
information we had. What feature in the data then implies that we are looking
at a KT transition broadened by  finite-size effects? Figure~\ref{fig:dvdi}
shows the same data as $d\ln E/d\ln J$ plotted against $J$ in log scale.  
Below the
KT transition in the low-temperature phase the slope $d\ln E/d\ln J$ for
smaller $J$ should be a constant given by $z(T)+1$. However, because of the
finite-size effect, the slope instead crosses over to 1 for smaller $J$. This
crossover produces a maximum for the curves in Fig.~\ref{fig:dvdi}.  Since the
crossover is a finite-size effect this means that the position of the maximum
is controlled by the variable $JL$ up to additional $T$-dependencies.  If the
additional $T$-dependencies are weak this implies that the maximum is occurring
at roughly the same $I (= JL)$ for all $T$ below the KT transition. This turns
out to be true for the 2D RSJ model as indicated by the full drawn lines
connecting the maxima in Fig.~\ref{fig:dvdi}. Above the KT transition the
maximum is instead controlled by the variable $J\xi$ and since the correlation
length $\xi$ strongly decreases with increasing $T$ above the KT transition
this means that the maximum above the KT transition should move to larger $J$
with increasing $T$. The crossover between these two behaviors should occur
near the temperature for which $L \approx \xi$, i.e., a $T$ somewhat larger
than $T_{KT}$. In Fig.~\ref{fig:dvdi} such a crossover occurs between $T=1.00$
and $1.10$ which is consistent with $T_{KT} \approx 0.90$. The inset in
Fig.~\ref{fig:dvdi} compares these maxima (denoted as 'max') with the values
$z(T)+1$ expected in the thermodynamic limit: the one obtained from a static
calculation [see Eq.~(\ref{eq:z})] (denoted as 'static') and the other one from
the scaling form $R \sim L^{-z(T)}$ (denoted as 'scale', data from
Ref.~\onlinecite{beom:big}) [see Eq.~(\ref{eq:ourscale}))]. As seen from the
inset, the maxima in Fig.~\ref{fig:dvdi} agree very well with the actual values
for the thermodynamic limit. This implies that the KT transition occurs close
to the temperature for which the maximum is 3, which according to the inset
gives a value close to $T=0.90$. This suggests that the signature of a KT
transition for a finite-size system is reflected in the behavior of the
positions of the maxima and that the transition is close to the $T$ for which
the maximum is 3.  
 
\begin{figure}
\centering{\resizebox*{!}{6.0cm}{\includegraphics{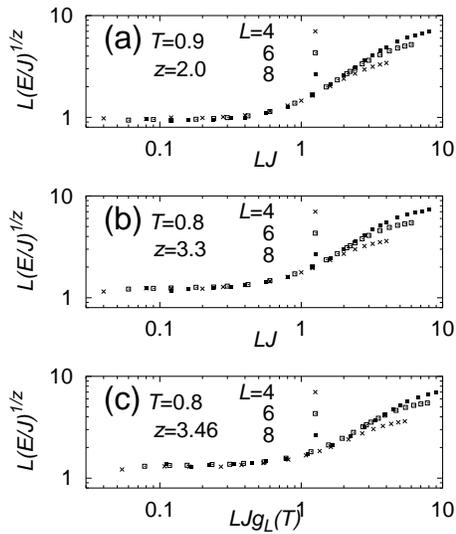}}}
\vskip 0.5cm
\caption{The size scaling given by Eq.~(\ref{eq:ourscale}) for the 2D RSJ model.
$L(E/J)^{1/z(T)}$ is plotted against $LJ$ for a fixed $T$ and the
three different sizes $L=4,6$, and 8.
The exponent $z$ is determined from a scaling collapse of the data:
(a) is for $T=0.90$ and (b) for $T=0.80$. In (a) and (b) a possible
additional
small $L$-dependence in Eq.~(\ref{eq:ourscale}) coming from $g_L$ is
ignored
and the values $z\approx 2$ and 3.3 are obtained for $T=0.90$ and
0.80, respectively.
In (c) the $L$-dependence from $g_L$ is included for $T=0.80$ and 
a good data collapse is obtained for the expected value $z=1/\epsilon T^{CG}-2\approx 3.46$ (see text).
}
\label{fig:Lscale}
\end{figure}

Next we demonstrate how the usual finite-size scaling given by
Eq.~(\ref{eq:ourscale}) works for the case of the 2D RSJ model. First we assume that
the $L$ dependence of $g_L(T)$ is weak, so that the scaling variable is to a good
approximation $JL$. Figure~\ref{fig:Lscale}(a) then verifies the
expected result that $z(T_{KT})=2$.~\cite{teitel,melwyn} This is done by
calculating the $I$-$V$ characteristics for the known KT transition temperature
$T_{KT}\approx 0.90$ of the 2D $XY$ model for some lattice sizes and then
plotting $L(E/J)^{1/z}$ against $LJ$. The $z$ value is
then determined by finding the best data collapse for small values of
$JL$. Since the low-temperature phase is a line of fixed points,
the same procedure can be used to determine $z(T)$ for any $T$ in the 
low-temperature phase.  Figure~\ref{fig:Lscale}(b) gives the same scaling
determination at $T=0.80$ in which case the value $z(T)=3.3$ which should be
compared to the value $z(T)=3.46$ obtained from the direct static calculation
of $z(T)=1/\epsilon T^{CG}-1$. The determination in Fig.~\ref{fig:Lscale}(c)
includes the $L$ dependence of $g_L(T)$ [the determination of $g_L(T)$ will be
given below] and with this included a good collapse is obtained for the value
$z(T)=1/\epsilon T^{CG}-1$. Figure~\ref{fig:Lscale} thus illustrates that the
exponent $z(T)$ can be determined from the finite-size scaling, depends on
temperature, and has the expected value $z(T)=1/\epsilon T^{CG}-1$.

\begin{figure}
\centering{\resizebox*{!}{5.5cm}{\includegraphics{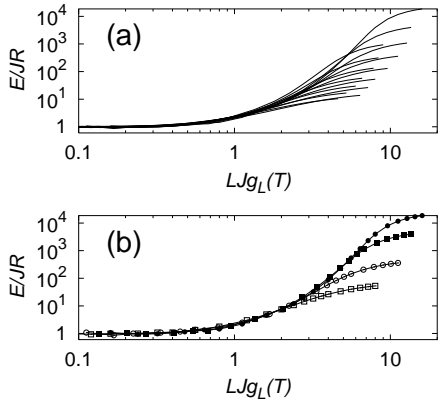}}}
\vskip 0.5cm
\caption{Demonstration of the existence of a scaling in the form of
Eq.~(\ref{eq:newscale}) for the size-dependent part of the $I$-$V$ characteristics
for the 2D RSJ model.  $E/JR$ is plotted against $LJg_L(T)$ and $g_L(T)$ is
determined from the condition of a data collapse. In (a) $g_L(T)$ is determined
for a fixed $L$ for the two sizes $L=6$ and 8 within the temperature intervals
$0.7\leq T\leq 1$ and $0.65\leq T\leq 1$, respectively.  A good data collapse
is obtained towards smaller values of $LJg_L(T)$ and furthermore the data for
$L=6$ and 8 also fall on top as required by Eq.~(\ref{eq:newscale}).  In
more detail, in (b) the same scaling plot is demonstrated 
for $T=0.9$ (open squares), 0.8 (open circles), 0.7 (filled squares) and 
0.65 (filled squares) in case of $L=8$. 
}
\label{fig:Tscale}
\end{figure}

As discussed in Sec.~\ref{sec:scaling}, the existence of the standard finite-size
scaling, as illustrated in Fig.~\ref{fig:Lscale}, does not rule out the
existence of a scaling of the form Eq.~(\ref{eq:newscale}) for small values of
$LJg_L(T)$. This scaling form implies that $E/JR$ is a function
only of $LJg_L(T)$. In Fig.~\ref{fig:Tscale} $E/JR$ is plotted
against $LJg_L(T)$ and shows that a function $g_L(T)$, which makes
the curves collapse, can indeed be found. The function $g_L(T)$ determined from
the two sizes $L=6$ and $8$ are shown in Fig.~\ref{fig:gt_R} and in 
Fig.~\ref{fig:z_g} $g_L(T)$ for $L=8$ is plotted against $z(T)$. 
We emphasize that the existence of a function $g_L(T)$,  which makes the 
data collapse, is by no
means trivial and that as a consequence the functions are well defined.
We again stress that the scaling in Fig.~\ref{fig:Tscale} [with $g_L(T)$ in
Fig.~\ref{fig:gt_R}] is compatible with the usual finite-size scaling
given by Eq.~(\ref{eq:ourscale}). 

As discussed in Sec.~\ref{sec:scaling} the Pierson scaling analysis, which
presumes a thermodynamic limit, is completely incongruent with the finite-size
scaling of the 2D $XY$ model discussed here. Nevertheless a connection from a
pragmatic standpoint would arise provided $g_L(T)=A_LR^{-\alpha}$, as also
discussed in Sec.~\ref{sec:scaling}. Figure~\ref{fig:gt_R} demonstrates that such a
relation does exist over a limited $T$ region. The exponent $\alpha\approx 1/6$
is found both for lattice sizes $6$ and $8$ and is well determined since $R$ has
a very strong $T$ dependence. As discussed in Sec.~\ref{sec:scaling} in connection
with Eq.~(\ref{eq:g}), the relation $g_L(T)=A_LR^{-\alpha}$, if it applied all
the way down to the temperature for which $z(T)=6$, would imply that $R$ vanishes at this $T$. This is,
from our finite-size perspective, the connection to the Pierson scaling method
and the putative thermodynamic transition. We illustrate this in
Fig.~\ref{pierson_rsj}(a) where the data for which the relation $g_L(T)=A_LR^{-\alpha}$
holds is plotted in the Pierson form using the $T^{CG}=0.125$ for which
$z(T)=1/T^{CG}-2=6$ which is where the would-be transition should occur. The
connection is that $1/z$ in Fig.~\ref{pierson_rsj}(a) is our $\alpha$ and that the
variable $\xi$ corresponds to $R^{-1/z}$. Furthermore $R^{-1/z}$ is assumed to
have the form $\exp(b/\sqrt{T^{CG}-T^{CG}_c})$ where $T^{CG}_c$ is where the
would-be transition should occur. As seen from the inset, $R$ can be
represented by the suggested form with the ``correct'' $T^{CG}_c=0.125$ and
a good scaling fit is obtained for the value $z=1/\alpha=6$. Again we stress
that there is no real phase transition or a vanishing resistance at this
$T^{CG}_c$; the resistance is caused by the finite size of the system and the
actual scaling property is only present over a limited range.  
Indeed when the lower temperatures $T=0.70$ and 0.65 are included,
a Pierson scaling plot with $\alpha \approx 1/6$ can no longer be
obtained. However, a Pierson scaling plot can again be obtained
if $\alpha$ is changed to $\alpha \approx 1/12$ and the two highest
temperatures are excluded as shown in Fig.~\ref{pierson_rsj}(b). 
This suggests that the Pierson scaling method has too much 
flexibility.~\cite{strachan}
 
\section{Re-analysis of Repaci's YBCO data} \label{sec:repaci}
In this section we compare our analysis of the 2D $XY$ model with the
measured $I$-$V$ characteristics for an ultrathin YBCO sample by 
Repaci {\it et al.}~\cite{repaci} Figure~\ref{repaci_iv} shows Repaci's $I$-$V$ data for $10{\rm
  K}\leq T\leq 40 {\rm K}$ (from left to right) which should
be compared to Fig.~\ref{fig:rsjall} for the 2D RSJ model. The thick lines have the
slopes one, three, and seven (from left to right), 
respectively, where the slope one represents
the linear resistive tail, slope three is the value for the KT transition 
and slope seven the value where the scaling over a limited region for
the 2D RSJ model would place a nonexistent ``ghost'' transition.
Repaci {\it et al.} attributed the linear resistance below the KT transition
slope three with free vortices induced by the finite size of the sample
(width 200 $\mu$m). The similarity between Fig.~\ref{fig:rsjall} and 
Fig.~\ref{repaci_iv} is in
accord with this interpretation. Figure~\ref{repaci_max} shows the Repaci data plotted
in the same way as the data for the 2D RSJ model in Fig.~\ref{fig:dvdi}. The
similarity is again striking. As in Fig.~\ref{fig:dvdi} the maxima in Fig.~\ref{repaci_iv} occur at approximately the same $I$ for lower $T$ (maxima larger than 2).
This is consistent with the finite-size interpretation given in
connection with Fig.~\ref{fig:dvdi} that the
position of the maxima is roughly determined by
$JL$. Consequently, when the size $L$ is the dominant length, the
maxima occur at the same $I$. However, for higher $T$ (maxima lower
than 2 in Fig.~\ref{repaci_max}) the correlation length $\xi$ is the dominant length ($\xi<L$)
and, since $\xi$ rapidly decreases with increasing $T$, the maxima
are displaced towards higher $I$. In parallel with Fig.~\ref{fig:dvdi} the
thermodynamic KT transition is expected to occur close to the maximum
value 3 which means $T\approx 27$K (compare full drawn line with slope
3 in Fig.~\ref{fig:rsjall}). Consequently $\xi$ is expected to become the dominant length for a
somewhat higher $T$ and this is consistent with the fact that the
maxima seem to start to move roughly around maximum value 2. The similarities between
Figs.~\ref{fig:rsjall} and \ref{fig:dvdi} for the 2D RSJ model and Figs.~\ref{repaci_iv} and \ref{repaci_max} for the Repaci data
support the interpretation of the Repaci data in terms of a finite-size
induced resistance below the thermodynamic KT transitions.

\begin{figure}
\centering{\resizebox*{!}{5.5cm}{\includegraphics{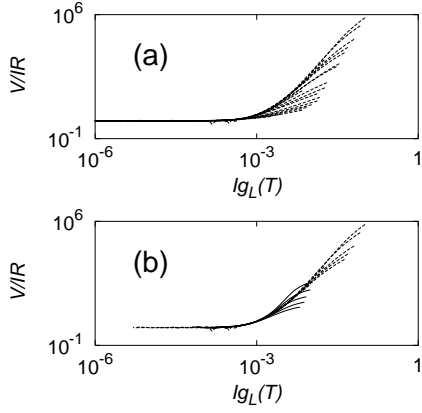}}}
\vskip 0.5cm
\caption{ Demonstration of the existence of a scaling in the form of
Eq.~(\ref{eq:newscale}) for Repaci's $I$-$V$ data.  $V/IR$ is plotted against
$Ig_L(T)$ and $g_L(T)$ is determined from the condition of a data collapse: (a)
shows that a good scaling collapse is obtained for the data in the interval
23K$\leq T\leq$40K (the highest $T$ corresponds to the lowest curve). Note the
striking similarity with the 2D RSJ data in Fig.~\ref{fig:Tscale}(a). 
This similarity is further
emphasized in (b) where the scaling function for the 2D RSJ model 
(full lines, note that $E/J=V/I$) is directly compared with the scaling function for
the Repaci data (dotted lines). The two scaling function appears to fall on to 
each other when allowing for a constant shift along the $x$ axis, suggesting
that they are closely related.
}

\label{repaci_scale}
\end{figure}

The next question is then whether the size scaling Eq.~(\ref{eq:newscale})
for the 2D RSJ model likewise applies to Recapi's YBCO data.
Figure~\ref{repaci_scale}(a) corresponds to Fig.~\ref{fig:Tscale}(a) for the RSJ model. The data in Fig.~\ref{repaci_scale}(a)
represents the data in the interval for which a linear resistance
$V/I=R$ is clearly present in the measurements (which means $23 {\rm
  K}\leq T\leq 40 {\rm K}$). The data collapse 
in Fig.~\ref{repaci_scale}(a) shows that $V/IR$ is only a
function of the scaling variable $JLg_L(T)$ when 
$JLg_L(T)$ is small enough, while Fig.~\ref{repaci_scale}(b) shows that the form of the
obtained scaling function is very similar to the one found for the 2D
RSJ model. This suggests that the cause of the scaling is the same
for the two cases, supporting the finite-size interpretation. The obtained scaling function $g_L(T)$ is plotted
in Fig.~\ref{repaci_gl}, and, just as in the 2D RSJ case, it is very well represented
by the $AR^{-\alpha}$ with $\alpha=1/6$. The range over which this
is true for the 2D RSJ model is roughly $2\leq z(T)\leq 3.5$ (see Fig.~\ref{fig:z_g}) and the
$g_L(T)$ in Fig.~\ref{repaci_gl} in fact covers the somewhat larger range 
$2\leq z(T)\leq 4.5$. 
As explained in Sec.~\ref{sec:rsj}, it is the fact that $g_L(T)$ is well represented by
$AR^{-\alpha}$ with $\alpha=1/6$ over a range around the KT transition which singles
out the temperature where $z(T)\approx 6$ as the ``ghost''
transition.
The temperature corresponding to the slope $z(T) + 1 =7$ is given by the
steepest full drawn line in Fig.~\ref{repaci_iv}.
However, in analogy with the 2D RSJ
model, we infer that there is in reality no vanishing resistance
except at $T=0$. In this sense it represents a ``ghost'' transition.

Since $g_L(T)$, $\alpha$, and $z(T)$ (estimated from the maxima in
Fig.~\ref{repaci_max}) are known, we can directly test Eq.~(\ref{eq:g}) by plotting
$\ln g_L(T)$ as a function of $\alpha z(T)/[1-\alpha z(T)]$.
According to the discussion of Eq.~(\ref{eq:g}) this should roughly be a
straight line. As seen from Fig.~\ref{repaci_gz} this is indeed the case for $T$
below $z \approx 2$ (marked by a vertical dashed line in Fig.~\ref{repaci_gz}, 
note that $z\approx 2$ is approximately where the finite-size effect sets in
according to Figs.~\ref{fig:dvdi} and ~\ref{repaci_max}).
If it
continued to be a straight line for lower $T$ then the condition
$1-\alpha z(T)=0$ implies $g_L=\infty$ and $R=0$ signaling the ``ghost''
transition for $z(T)=1/\alpha$.

An alternative way of demonstrating the existence of the scaling
Eq.~(\ref{eq:newscale}) with the scaling variable $IR^{-\alpha}$ is to
note that this scaling implies that $V/IR$ for a given $IR^{-\alpha}$ is independent of $T$ (see Ref.~\onlinecite{haan}, where a similar method
has been used).
Thus we can pick a given $V/IR$ and mark the value $I$ and $RI$ for
which it occurs. As seen in Fig.~\ref{repaci_gz} this construction gives rise to a
straight line for a given value of $V/IR$
in the plot of $V$ against $I$ in log-log scale 
and the slope of such a line is
according to the scaling assumption
$1/\alpha+1$. As apparent from Fig.~\ref{expVI} this alternative test of the
scaling is also borne out and gives values in the vicinity of $1/\alpha\approx 6$.

Finally in Fig.~\ref{repaci_piersonscale} we plot the result from the Pierson scaling
method for comparison.

\section{Discussion and Conclusions} \label{sec:conc}
\begin{figure}
\centering{\resizebox*{!}{5.5cm}{\includegraphics{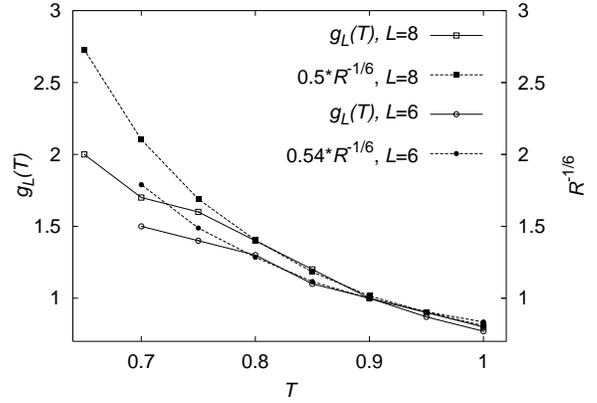}}}
\vskip 0.5cm
\caption{The function $g_L(T)$ determined for the 2D RSJ model with
size $L=6$ (open circles) and 8 (open squares). It is shown that the
function $g_L$ for both sizes over a limited $T$ interval is well
approximated by $g_L(T)\propto R^{-\alpha}$ with $\alpha\approx 1/6$.   
}
\label{fig:gt_R}
\end{figure}

\begin{figure}
\centering{\resizebox*{!}{5.5cm}{\includegraphics{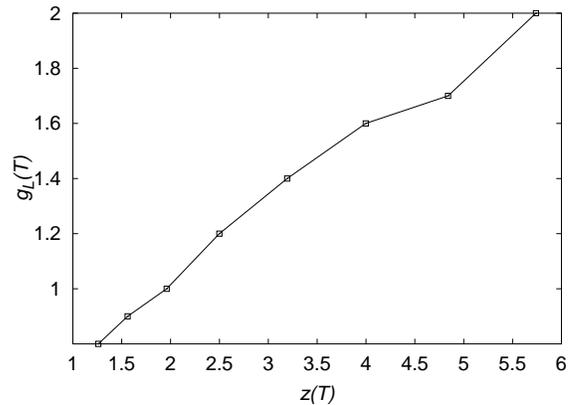}}}
\vskip 0.5cm
\caption{The function $g_L(T)$ as a function of $z(T)$ for the 2D
RSJ model with size $L=8$. The proportionality between $g_L(T)$
and $R^{-1/6}$ holds in the interval $1.3<z(T)<3.5$ (see Fig.~\ref{fig:gt_R}).   
}
\label{fig:z_g}
\end{figure}

\begin{figure}
\centering{\resizebox*{!}{5.5cm}{\includegraphics{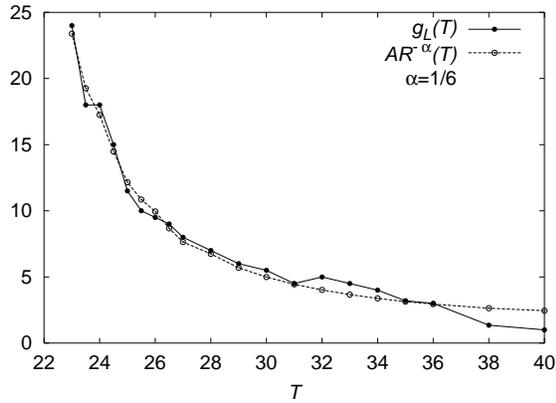}}}
\vskip 0.5cm
\caption{The function $g_L(T)$ determined for Repaci's YBCO data: The filled
circles give the values of $g_L(T)$ determined from the data collapse in
Fig.~\ref{repaci_scale}(a). The open circles demonstrates that the obtained $g_L(T)$ is well
represented by $g_L(T)=AR^{-\alpha}$ with $\alpha=1/6$.   Compare Fig.~\ref{fig:gt_R} for the
2D RSJ model.  
}
\label{repaci_gl}
\end{figure}

\begin{figure}
\centering{\resizebox*{!}{5.5cm}{\includegraphics{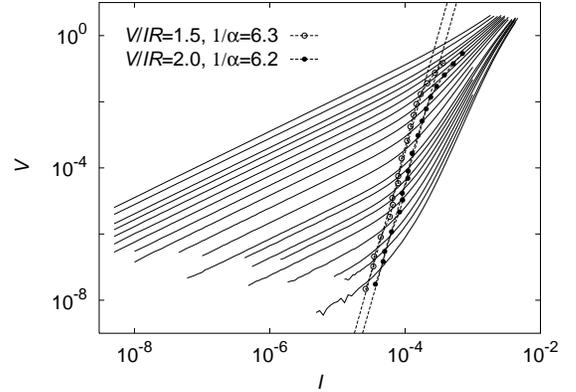}}}
\vskip 0.5cm
\caption{Direct determination of the exponent $\alpha$ for Repaci's YBCO data.
Eq.~(\ref{eq:newscale}) together with $g_L\propto R^{-\alpha}$ means that $V/IR$ is
a function of $IR^{-\alpha}$. $V (=RI)$ is plotted 
as a function of $I$ for a
given fixed value of $V/IR=$const. (open and filled circles correspond to
const.=1.5 and 2.0, respectively). Consequently, the scaling implies that the
data points should fall on straight lines given by $RI\propto I^{-1/\alpha -1}$
and the slope of these lines should be given by $-1/\alpha - 1$. As seen this
prediction is borne out and the obtained values are $\alpha^{-1}=6.3$ and 6.2
in close agreement with the value $\alpha^{-1}\approx 6$ obtained from 
Fig.~\ref{repaci_gl}.  
}
\label{expVI}
\end{figure}

We have shown from simulations that the finite-size induced features of the 
$I$-$V$
characteristics for the finite-size 2D RSJ model obey a scaling of the form given by
Eq.~(\ref{eq:newscale}) over a limited temperature range in the
vicinity of the KT transition. The $T$-dependence of this
scaling is absorbed in a function $g_L(T)$ for each fixed size $L$. Furthermore it was shown
that this function $g_L(T)$ over a limited region in the vicinity of
the KT transition is to a good approximation proportional to
$R^{-\alpha}$. For the two sizes we studied ($L=6$ and 8) we found
that $g_L(T)$ was different, yet the exponent $\alpha\approx 1/6$ was
found in both cases. We noted that if the scaling form
Eq.~(\ref{eq:newscale}) was valid down to low enough $T$ and that at the
same time the proportionality $g_L(T)\propto R^{-\alpha}$ was valid to
low enough $T$, then this would imply that the resistance vanishes at
the $T$ for which $z(T)=1/\alpha$.
However, we stressed that in practice there is no such transition
for the finite-size 2D RSJ model and introduced the term ``ghost''
transition for the transition which is not there.
The fact that the transition does not occur is linked to
the fact that the scaling in practice breaks down for lower $T$,
as clearly shown in Fig.~\ref{fig:gt_R}.
For the finite-size 2D RSJ model the resistance vanishes only for $T=0$.

\begin{figure}
\centering{\resizebox*{!}{5.5cm}{\includegraphics{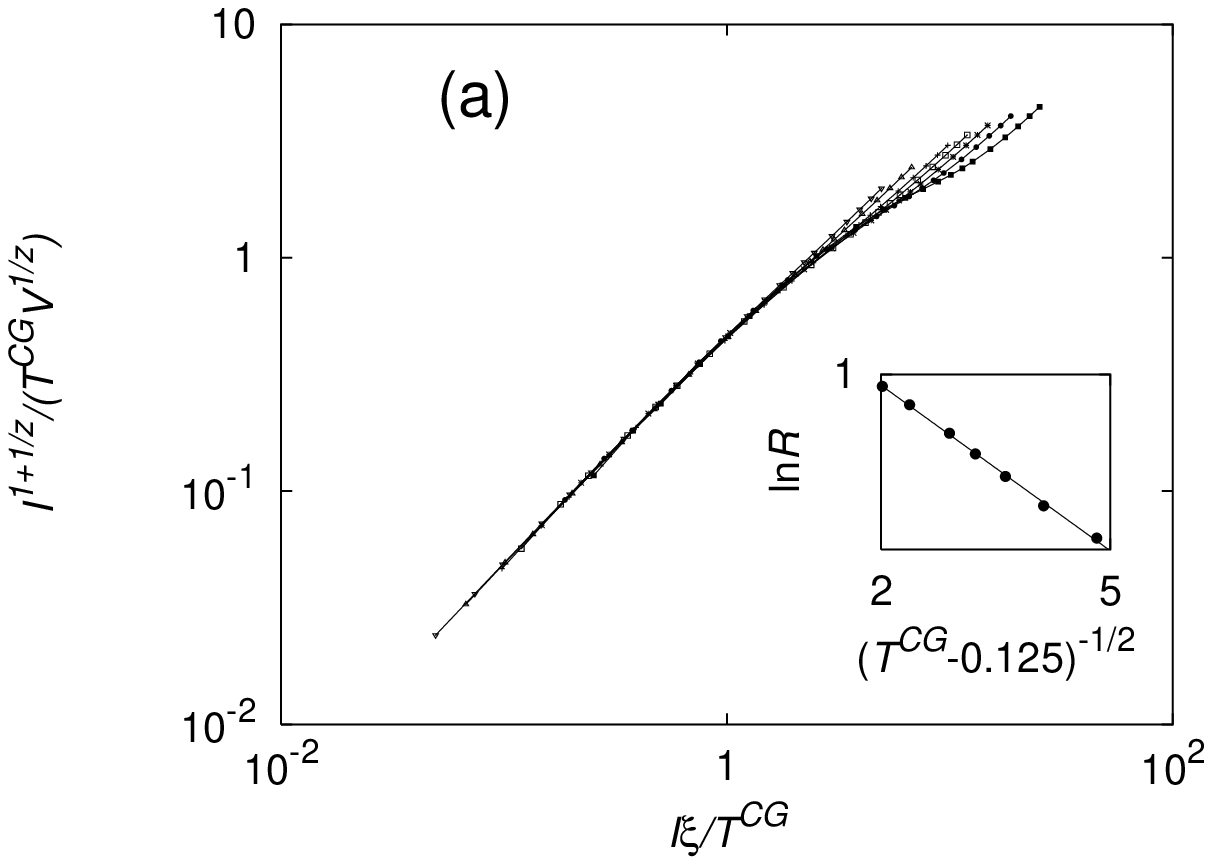}}}
\centering{\resizebox*{!}{5.5cm}{\includegraphics{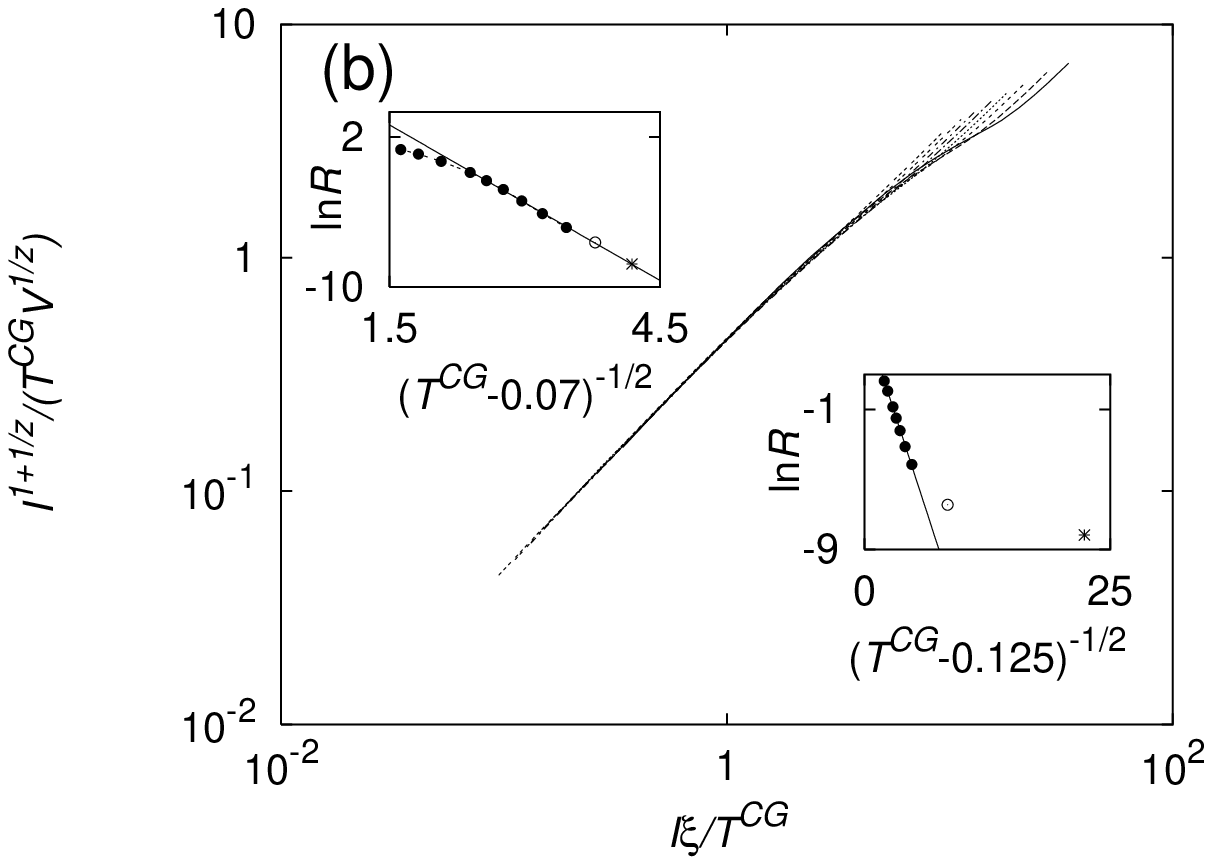}}}
\vskip 0.5cm
\caption{(a) Demonstration that the scaling obeyed by the finite-size induced part
of the $I$-$V$ characteristics for the 2D RSJ model with size $L=8$ can be recast
into the Pierson form: $I^{1+\alpha}/(T^{CG}V^{\alpha})$ is plotted against
$IR^{-\alpha}/T^{CG}$ and a good scaling collapse is found with $\alpha=1/6$
for the data in the interval $0.80\leq T \leq 1.2$.  Identifying $1/\alpha$
with $z$ and $R^{-\alpha}$ with $\xi$ gives the connection $R^{-1/z}\propto \xi$
and makes the scaling look like the Pierson form. The inset shows that $\ln
R\propto 1/\sqrt{T^{CG}-0.125}$ suggesting that the resistance would vanish
at $T^{CG}=0.125$ and at this $T$ ($\approx 0.64$) the dynamic critical exponent $z(T)$ is to
a good approximation given by $z(T)=1/T^{CG}-2=6$ consistent with the condition
$z(T)=1/\alpha$.  (b) Pierson scaling plot including the lower temperatures
$T=0.70$ (open circles) and 0.65 (asterisks) for which 
$g_L \sim R^{-1/6}$ fails (see Fig.~\ref{fig:gt_R}). As seen in the upper left inset the same
approximation for $R$ can be used provided $\sqrt{T^{CG} - 0.125}$ is
replaced by $\sqrt{T^{CG} - 0.07}$. The lower right inset shows
that   $\sqrt{T^{CG} - 0.125}$ fails for the lowest values of $T$.
In the main part, 
it is shown that a good scaling plot can still be obtained provided the
two highest temperatures are excluded but the ensuing $z$ value is
now $z\approx 12$. This illustrates that the Pierson construction has
too much flexibility.~\cite{strachan}
}

\label{pierson_rsj}
\end{figure}

We also stressed that the theoretical basis for the Pierson scaling method is
incompatible with the finite-size effect we study for the 2D RSJ model.
However, in practice the Pierson method assumes that for the high-temperature
branch of the scaling form Eq.~(\ref{eq:p1}) one has $\xi\propto R^{-\alpha}$.
Thus the high-temperature branch of the Pierson scaling is, from a pragmatic
point of view, connected to our finite-size scaling. We explicitly demonstrated
this connection in Fig.~\ref{pierson_rsj} for the finite-size 2D RSJ model.

\begin{figure}
\centering{\resizebox*{!}{5.5cm}{\includegraphics{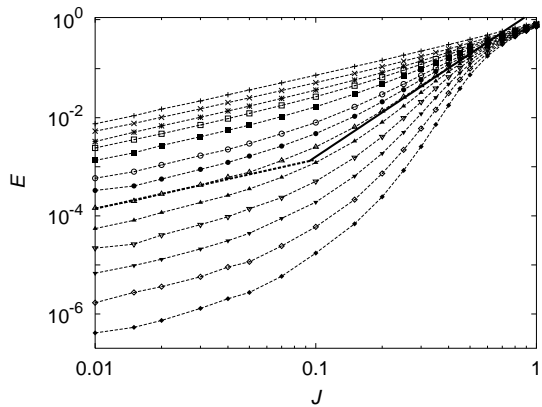}}}
\vskip 0.5cm
\caption{$I$-$V$ characteristics for the 2D RSJ model on a square lattice
with the finite size
$L=8$ plotted as $E=V/L$ against $J=I/L$ in
log scales at temperatures $T=2.0$, 1.5, 1.3, 1.2, 1.1, 1.0, 0.95, 0.90,
 0.85, 0.80, 0.75, 0.70, and 0.65 (from top to bottom).
The thick solid line corresponds to 
$V\propto I^3$ and has to a good approximation the same slope as the
$T=0.90$ curve over a finite-current region. For smaller currents the
system becomes Ohmic, $E=RJ$, due to the finite size, which
corresponds to the thick broken line.  
}
\label{fig:rsjall}
\end{figure}

\begin{figure}
\centering{\resizebox*{!}{5.5cm}{\includegraphics{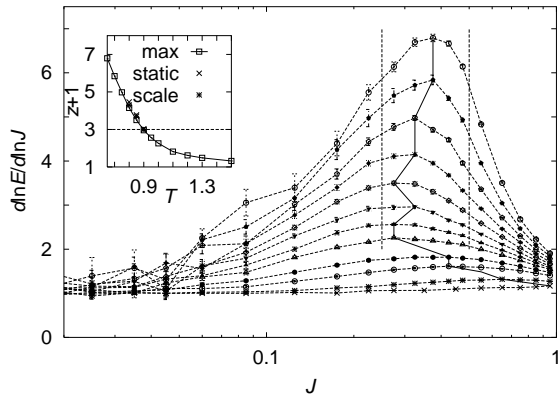}}}
\vskip 0.5cm
\caption{The $I$-$V$ data for the 2D RSJ model with $L=8$ plotted as
$d\ln E/d\ln J$ against $J$ in log scale. From bottom to top the
curves correspond to $T = 2.0$, 1.5, 1.3, 1.2, 1.1, 1.0, 0.95, 0.90,
0.85, 0.80, 0.75, 0.70, and 0.65. The full drawn line segments connect the
maxima.
Down to $d\ln E/d\ln J\approx 2$ the maxima occur at about the
same $J$ (between the vertical broken lines) and then move to higher
$J$.
The inset illustrates that the maxima of $d\ln E/d\ln J$
to a good approximation are given by $z+1$ where $z$ is determined in
two different ways (see text). The horizontal broken line in the inset
corresponds to the KT value $a=z+1=3$ and the crossing point gives the
estimate $T_{KT}\approx 0.9$.}
\label{fig:dvdi}
\end{figure}

\begin{figure}
\centering{\resizebox*{!}{5.5cm}{\includegraphics{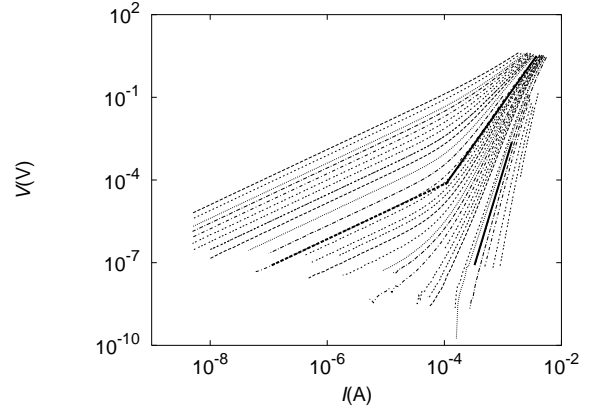}}}
\vskip 0.5cm
\caption{The $I$-$V$ characteristics for Repaci's YBCO data: $V$ is plotted as a
function of $I$ for fixed $T$ where 10K$\leq T\leq$40K 
(from left to right).  
The thick full line in the middle has the slope 3, which is the KT
transition value and corresponds to the steepest slope of the data for $T=27$K.
The right thick full line has the slope 7, which corresponds to the
"ghost" transition for the finite-size 2D RSJ model and which is the steepest
slope of the data for $T=18$K.  The thick broken line has the slope 1 and
corresponds to Ohmic resistance $V=RI$. Note the striking similarity with the
2D RSJ data in Fig.~\ref{fig:rsjall}.   
}
\label{repaci_iv}
\end{figure}

\begin{figure}
\centering{\resizebox*{!}{5.5cm}{\includegraphics{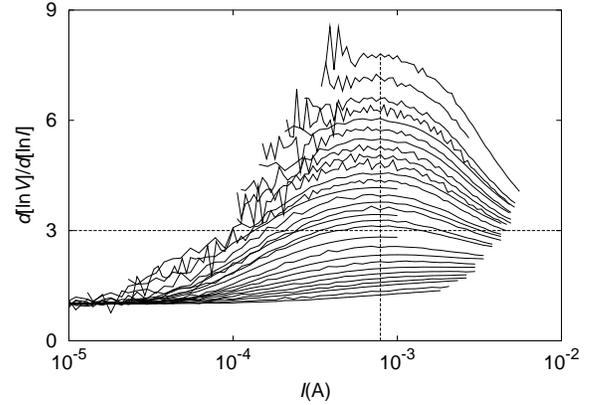}}}
\vskip 0.5cm
\caption{ The $I$-$V$ data for Repaci's YBCO data plotted as $d\ln V/d\ln I$
against $I$ in log scale for 17K$\leq T\leq$40K.  The values of
the maxima increase with decreasing $T$ and give estimates of $z(T)+1$.
The horizontal broken line corresponds to the KT value $z(T)+1=3$.  The
vertical broken line gives the approximate position for the maxima when
$d\ln V/d\ln I \protect\gtrsim 2$, showing that the maxima occur at
approximately the same $I$. 
The maxima for $d\ln V/d\ln I \protect\lesssim 2$ occur
at higher $I$ with increasing $T$.  Note the striking similarity
with Fig.~\ref{fig:dvdi} for the finite-size 2D RSJ model.
}
\label{repaci_max}
\end{figure}

\begin{figure}
\centering{\resizebox*{!}{5.5cm}{\includegraphics{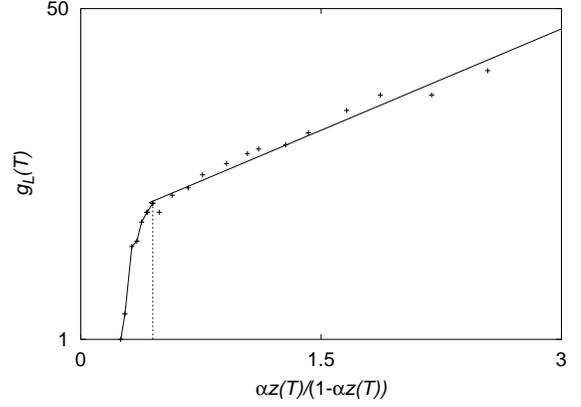}}}
\vskip 0.5cm
\caption{Repaci's data analyzed according to Eq.~(\ref{eq:g}): $g_L(T)$ in log scale is
plotted against $\alpha z(T)/[1-\alpha z(T)]$ where $\alpha$ according to
Fig.~\ref{repaci_gl} is 1/6 to a good approximation and $z(T)$ is determined from the maxima
in Fig.~\ref{repaci_max}. As seen, the data fall approximately along a straight line from
about where $z(T)=2$ (vertical broken line). This corresponds to the $T$ region
where the maxima in Fig.~\ref{repaci_max} occur for the same current $I$ suggesting that this
is the finite-size dominated part of the $I$-$V$ characteristics. If the data
would continue to fall on the straight line all the way to infinity, this would
imply $g_L(T)=\infty$ and equivalently a vanishing resistance for
$z(T)=1/\alpha$. Since this does not happen in practice we use the term
``ghost'' transition. 
}
\label{repaci_gz}
\end{figure}

We compared the data obtained for the finite-size 2D RSJ model with
the YBCO data by Repaci {\it et al}.~\cite{repaci} From the strong similarity
manifested by Figs.~\ref{fig:rsjall}, \ref{fig:dvdi}, \ref{repaci_iv}, 
and \ref{repaci_max} we concluded that Recapi's
data correspond to a KT transition around $27$ K and that the
resistance below this $T$ is finite-size induced. This is in accord
with the original conclusion by Repaci {\it et al.}~\cite{repaci} Here we want
to emphasize that finite size essentially means a cutoff in the
logarithmic vortex interaction. Apart from the actual finite size of
the system this can be caused by a small remnant magnetic field or by
a finite perpendicular penetration depth.~\cite{minnhagen:rev}   

We demonstrated the existence of a scaling of the form
Eq.~(\ref{eq:newscale}) also for Repaci's YBCO data and furthermore
demonstrated that the shape of the scaling function appeared to be the
same as for the finite-size 2D RSJ model. This we interpreted as that the
scaling for the YBCO data and for the 2D RSJ model have a common
origin. This further supports the finite-size interpretation of the
YBCO data. The function $g_L(T)$ was just as in the 2D RSJ model well
represented by $g_L(T)\propto R^{-\alpha}$ with the same exponent
$\alpha\approx 1/6$ as for the 2D RSJ model. The remarkable similarity
between the results for the finite-size 2D RSJ model and Repaci's
YBCO data clearly suggests that the same physics is reflected in the
two systems. Our conclusion is then that, since it reflects a finite-size
induced feature for the 2D RSJ model, it also reflects a finite-size induced feature for the
YBCO data. Consequently, we conclude that the transition to a vanishing
resistance implied by the condition $z(T)=1/\alpha=6$ is also for the
YBCO data a ``ghost'' transition.

\begin{figure}
\centering{\resizebox*{!}{5.5cm}{\includegraphics{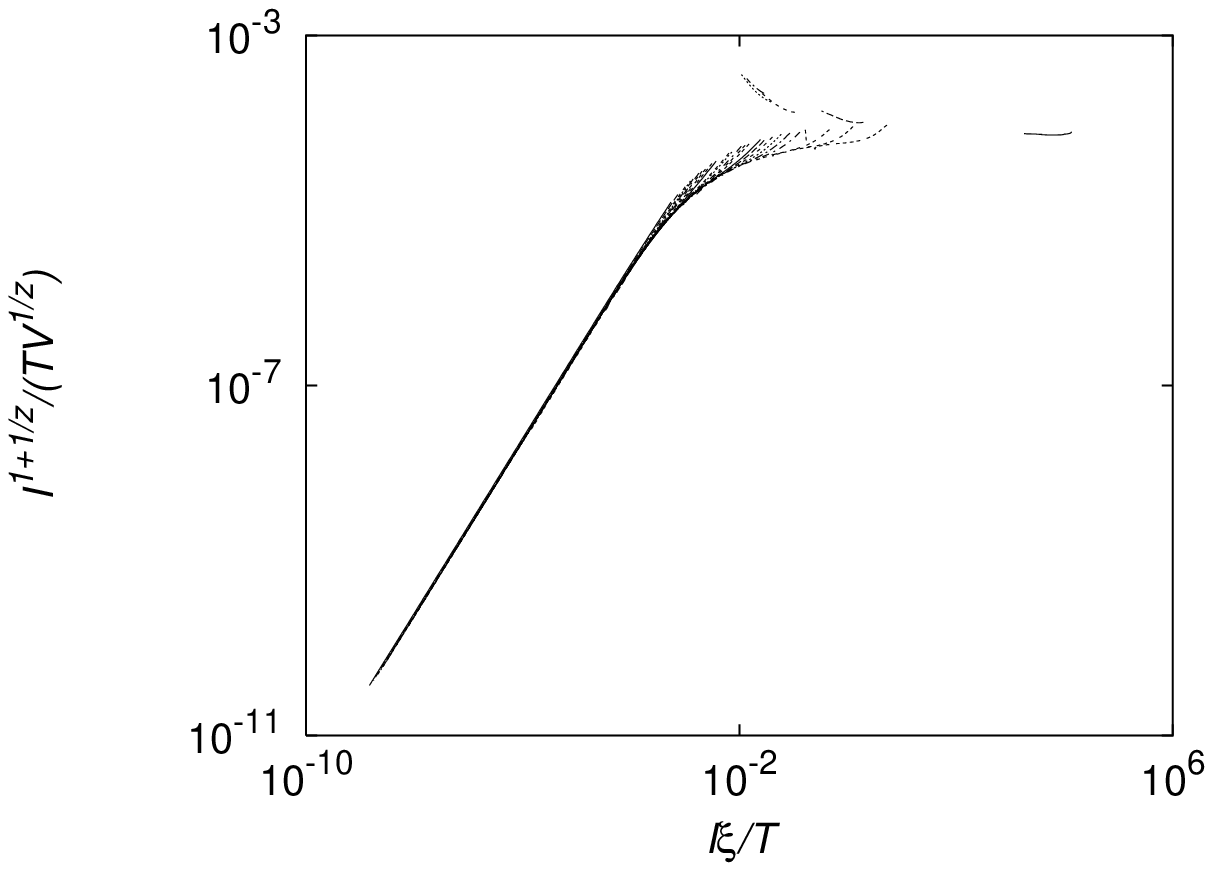}}}
\vskip 0.5cm
\caption{The Pierson scaling plot for Repaci's YBCO data:~\cite{pierson}
$I^{1+1/z}/(TV^{1/z})$ is plotted against $I\xi/T$ where it is assumed that
$\xi\propto R^{-1/z}$. The data is plotted in log scales and the construction
makes the scales very extended which exaggerates the goodness of the data
collapse. The high-$T$ branch of the scaling plot corresponds to 18K$\leq
T\leq$40K and is given by the lower scaling curve and should be compared to the
scaling of the finite-size 2D RSJ model in Fig.~\ref{pierson_rsj}. The low-$T$ branch contains
data for 10K$\leq T\leq$16K. A transition of a vanishing resistance at $T=17$K
with $z=5.9$ was concluded from this scaling construction.~\cite{pierson} We
claim that this scaling plot does not reflect any real transition to a
vanishing resistance.  }
\label{repaci_piersonscale}
\end{figure}

Pierson {\it et al.} in Ref.~\onlinecite{pierson}, when analyzing the Repaci's
YBCO data with the Pierson scaling method, inferred a transition
to a vanishing resistance for $z(T)=5.9$.
This is closely the same value as where our finite-size scaling places
the "ghost" transition both for the same experimental data and for the 2D
RSJ model.
This we believe is because the high-temperature branch of the Pierson
scaling
from a pragmatic point reflects the same behavior as our finite-size
scaling.
The low-temperature branch of the Pierson scaling is as seen in 
Fig.~\ref{repaci_piersonscale}
less convincing and it does not have any relation to our finite-size scaling. 
The fact that Pierson {\it et al.} for very many cases have found that their
scaling
suggests a transition at $z(T)\approx 6$ from our perspective suggest
a common origin. The fact that the same result is consistent with the
finite-size 2D RSJ model suggests to us that this common origin is
linked
to the finite-size induced features of the $I$-$V$ characteristics in the vicinity of the KT transition.~\cite{footnew2}

Is the scaling property for the finite-size induced feature of the
$I$-$V$ characteristics
found in the present paper a real phenomena or just some accidental
coincidence? Such accidental coincidences can of course never be completely
ruled out.
However, our analysis really leaves little room for any such accidental coincidence.
Yet, we admit that we do not have a plausible explanation at the
moment.
Thus the enigma with the scaling and the "ghost" transition calls
for further theoretical, experimental and computational efforts. We
have in the present paper described how the scaling property is
reflected in the $I$-$V$ data from a practical point of view and how to
analyze the data from such a perspective.

Finally we stress that our interpretation in terms of a finite-size
induced feature of the
$I$-$V$ characteristics is entirely compatible with the established view
of the KT transition with $z=2$ and the usual finite-size scaling at a
fixed $T$ in the low-temperature phase,
as we explicitly showed for the 2D RSJ model. 
In our finite-size interpretation there is no
real phase transition but only a reflection of a "ghost" transition.
This also agrees with the view by Strachan {\it et al.} in
Ref.~\onlinecite{strachan}
that the apparent vanishing of the resistance in Recapi's data is not
real but an artifact of the finite-voltage threshold in the
measurements.
On the other hand our conclusions are entirely incompatible with the
interpretation
put forward by Pierson {\it et at.}~\cite{pierson} that there is a
real resistive transition and that this is a KT transition with
$z\approx 6$.
Yet our analysis from a pragmatic point of view supports the
original
finding by Pierson {\it et al.}~\cite{pierson} that there is after all an
intriguing scaling feature in the $I$-$V$ characteristics associated with
2D vortex fluctuations and that this scaling phenomena appears to be quite general. 

\acknowledgments
The research was supported by the Swedish Natural Science Research Council
through Contract No. FU 04040-322. The authors acknowledge M. Friesen, 
S.W. Pierson, D.R. Strachan, C.J. Lobb, and R.S. Newrock for useful discussions 
and the two latter for providing us the experimental data published in 
Ref.~\onlinecite{repaci}.

\narrowtext

\end{multicols}
\newpage

\widetext
\begin{multicols}{2}
\vskip 1.5cm

\end{multicols}
\narrowtext

\begin{thebibliography}{99}
\bibitem{foot:EJ} Throughout the present paper, we measure current-voltage
($I$-$V$) characteristics in terms of the electric
field $E$ and the current density $J$, which are given by
$V \equiv LE $ and $I \equiv LJ$ for a 2D $L\times L$ system.
\bibitem{footnew1} Below the KT transition the infinite correlation
length $\xi=\infty$ leads to the well-known non-linear IV-characteristics
$E\propto J^a$ where $a$ is an exponent larger than 3.
\bibitem{foot} At a temperature below the KT transition the system
is controlled by a critical point on the line of critical points 
within the low-temperature phase, and not by the one
describing the KT transiton.
\end{thebibliography}
\end{document}